\documentclass[pra,aps,twocolumn,showpacs]{revtex4}
\usepackage{amssymb,amsmath,amsthm,amsbsy,epsfig,color,graphicx,times,bbm}
\usepackage[ansinew]{inputenc}
\usepackage[english]{babel}
\usepackage{accents}
\usepackage{braket}
\usepackage{amsfonts}
\usepackage{booktabs}
\usepackage{tabularx}
\usepackage{array}
\usepackage{hyperref}
\hypersetup{pdfstartview=FitH}
\usepackage{latexsym}
\usepackage{amsthm}
\newtheorem{theorem}{Theorem}
\newcommand{\virg}[1]{``#1''}

\begin{document}

\title{Topological and nematic ordered phases in many-body cluster-Ising models}

\author{S. M. Giampaolo}
\affiliation{University of Vienna, Faculty of Physics, Boltzmanngasse 5, 1090 Vienna, Austria;\\
CNR-IOM DEMOCRITOS Simulation Center, Via Bonomea 265, I-34136 Trieste, Italy}

\author{B. C. Hiesmayr}
\affiliation{University of Vienna, Faculty of Physics, Boltzmanngasse 5, 1090 Vienna, Austria}

\begin{abstract}
We present a fully analytically solvable family of models with many-body cluster interaction and Ising interaction. This family exhibits two
phases, dubbed cluster and Ising phases, respectively. The critical point turns out to be independent of the cluster size $n+2$ and is reached
exactly when both interactions are equally weighted. For even $n$  we prove that the cluster phase corresponds to a nematic ordered phase and
in the case of odd $n$ to a symmetry protected topological ordered phase. Though complex, we are able to quantify the multi-particle entanglement
content of neighboring spins. We prove that there exists no bipartite or, in more detail, no $n+1$-partite entanglement. This is possible since the
non-trivial symmetries of the Hamiltonian restrict the state space. Indeed, only if the Ising interaction is strong enough (local) genuine
$n+2$-partite entanglement is built up. Due to their analytically solvableness the $n$-cluster-Ising models serve as a prototype for studying non
trivial-spin orderings and due to their peculiar entanglement properties they serve as a potential reference system for the performance of quantum information
tasks.
\end{abstract}


\maketitle

\section{Introduction}

In many-body systems described by classical mechanics the presence of an ordered phase is connected to the spontaneous breaking of symmetries
associated with local order parameters. A system consisting of classical spins, for instance, may admit a ground
state having all spins oriented along a given direction. Such ground states simultaneously break the spin-rotation and the time-reversal symmetry
witnessed by a non-vanishing magnetic moment.

Considering quantum systems, in contrast, one finds also different phases connected to some physical quantity but not necessarily to the magnetic
moment. The paradigmatic example is a translation invariant spin-$\frac{1}{2}$ chain for which the ground states correspond to the so called valence
bond states, i.e. states build up by tensor products of maximally entangled bipartite states~\cite{A1973,A1987}. In such systems neither the
spin-rotation nor the time-reversal symmetry is broken, nevertheless, it is possible to define order parameters characterizing the phases.
Typically nematic phases occur if at least one symmetry of the Hamiltonian is broken which are phases with long range ordering, i.e. defined by
order parameters on a finite set of sides. Other examples intensively discussed are topological order phases~\cite{W1990,KW2009} that, for instance,
are associated with the robustness of ground state degeneracies~\cite{WN1990}, are quantized non-Abelian geometric phases~\cite{W1990} or possess
patterns due to long-range quantum entanglement~\cite{CGW2010}.

Frustration occurs for systems with competing interactions or non-trivial geometries and can be related to quantum entanglement~\cite{GHI2014}.
Non-trivial spin orders are usually found if an interplay between frustration and quantum fluctuations is at work resulting in chiral, nematic or
general multipolar phases. In contrast to topological phases, even in the case of a vanishing magnetic moment the spin-rotation symmetry is
broken~\cite{AG1984,LMM2011}. These phases are also interesting from the point of applications. The topological ordered phases play
a fundamental role in the spin liquids~\cite{IHM2011,ZGV2011} and in non-Abelian fractional Hall systems~\cite{LH2008} and are
predicted to play a key role in the future development of fault-tolerant quantum computers~\cite{K2003}. The nematic order is usually found in
materials commercially used in the liquid crystal technology~\cite{TGC2010} such as LCDs (liquid crystal display).

Non-trivial ordered spins appear usually for higher dimensional systems (lattices) or sites with more than two degrees of freedom
(spins higher than $\frac{1}{2}$). Exceptions are the frustrated one dimensional ferromagnetic spin-$\frac{1}{2}$ chain in
an external magnetic field having a nematic ordered phase~\cite{C1991,HHV2006} and the one dimensional cluster-Ising model exhibiting a symmetry
protected topological ordered phase~\cite{SAFFFPV2011,MH2012,GH2014}.

In general, mathematical tools to handle such systems are rare and only few very specific Hamiltonians have been found to be analytically solvable.
The present paper introduces a huge class of analytically solvable one dimensional models with two degrees of freedoms (spin-$\frac{1}{2}$)
exhibiting both topologically and nematic ordered phases, which we dub $n$-cluster-Ising models. The index $n=1,2,\dots$ refers to the presence of
an $n+2$-body interaction, a cluster size of $n+2$. The physical systems under investigation are characterized by two competing interactions, a
two-body Ising interaction along the $y$-axis and an $n+2$-body interaction along the $x$-axis and the $z$-axis. The Hamiltonian of the family of
models can be written as
 \begin{equation}
  \label{Hamiltonian}
\!\!\! H^{(n)}\!\!=\!\!J\!\!\left(\!\sin(\phi)\!  \sum_{j} \!\sigma_{j}^y \sigma_{j+1}^y \!-\!\cos(\phi)\! \sum_{j}\! \sigma_{j}^x O^z_{j,n} \sigma_{j+n+1}^x
 \! \right)
 \end{equation}
where $J$ has the dimension of an energy (which we set equal to one in the computation) and $O^z_{j,n}$ stands for
\begin{equation}
\label{Oz_operator}
 O^z_{j,n} = \bigotimes_{k=1}^{n} \sigma_{j+k}^z \; .
\end{equation}
Via the parameter $\phi$ the relative weight of the two interactions is controlled: When $\phi$ approaches $0$ the system is dominated
by the multi-body interaction whereas when $\phi$ tends to $\pi/2$ the system is dominated by the (anti-ferromagnetic) Ising interaction.

We show that this family of models can be analytically solved (Sec.~\ref{Solution}) and how the spin correlation function can be obtained
(Sec.~\ref{Spincorrelation}). We prove that there is a quantum critical point at $\phi_c=\pi/4$  separating the cluster phase from the Ising phase.
This corresponds to the case when both interactions have equal weights. This critical point $\phi_c$, surprisingly, does not depend on the
$n+2$-body interaction, hence, it does not dependon the cluster size. In strong contrast to the relevant ordering in the cluster phase that depends strongly on $n$:
In case of odd $n$ a symmetry protected topologically ordered phase is present, whereas for even $n$ a nematic phase is present. In both cases we
determine the order parameter (string order parameter for the topological ordered phase and block order parameter for the nematic phase) as
well as the order parameter of the Ising phase (Sec.~\ref{Orderparameters}).

In the next step we study the various entanglement properties of the family of models (Sec.~\ref{Entanglementproperties}). The first observation is
that for any $n$ and $\phi$ -- as proven for the standard cluster-Ising model ($n=1$) in Ref.~\cite{GH2014} -- there is no bipartite entanglement.
Picking out any two spins the state is separable. Indeed, we find that this family of Hamiltonians lead to ground states that possess
\textit{genuine} $k=n+2$-partite entanglement between any contiguous spins and any $k<n+2$-partite entanglement vanishes. The symmetries in the
state space of the ground states force the reduced state of any $n+2$ adjacent spins into a so-called $X$-form~\cite{YE2007}, i.e. by applying
certain local unitary operators the reduced density matrix has only non-zero entries on the two diagonals. Due to this form we can exactly
evaluate a measure for genuine multipartite entanglement~\cite{HH2008,MCCSGH2011,HHBE2012}, i.e. quantify the entanglement content. So far, long
range multipartite entanglement close to a phase transition has been studied in terms of entanglement witnesses, e.g. for the $XXZ$ spin
chain~\cite{SRPCS2014} or for the $XY$ model~\cite{GH2013,Discord2014,CRGB2013}. Having this strong tool at hand, a measure of genuine multipartite
entanglement, we find that non-zero \textit{genuine} multipartite entanglement is only non-zero in the Ising phase $\phi>\phi_c$ (except $n=1$),
thus exhibiting a fortunate behaviour for applications such as utilizing these quantum systems for quantum algorithms.

The block entanglement properties are studied with focus around the quantum phase transition. Via the relation between conformal field
theory~\cite{HLW1994} and the divergence of the block entanglement at the quantum phase transition we are able to evaluate the central charges of
the models that turns out to depend on $n$.

Last but not least we conclude (Sec.~\ref{Conclusions}) by discussing the interplay between the characterization of the many-body systems by ordered
parameters and by symmetries in the Hilbert-Schmidt space of the ground states revealing the entanglement properties.

\section{Solution of the models}
\label{Solution}

In this section we present how to compute analytically the ground states of the models under investigation. The idea is to map the Hamiltonian,
Eq.~(\ref{Hamiltonian}), of spin-$\frac{1}{2}$ particles into non-interacting fermions moving freely along the chain only obeying Pauli's exclusion
principle. This method works even for the case in which the length of the system diverges~\cite{LSM1961,BM1971}. Having finally computed the energy
density function we find a phase transition that is further analysed in Sec.~\ref{Orderparameters}.

The mapping of a spin model to a fermionic one is obtained by applying the Jordan-Wigner transformation~\cite{JW1928}. Providing the correct
anti-commutation rules in the Jordan-Wigner transformation one associates the local spin operators with non-local fermionic operators
\begin{eqnarray}
\label{JWT}
 c_j = \bigotimes_{k=1}^{j-1}\left( \sigma^z_k \right) \sigma_j^- \;, \;\; \;& & \; \; \;
 c_j^\dagger = \bigotimes_{k=1}^{j-1} \left( \sigma^z_k \right) \sigma_j^+ \; ,
\end{eqnarray}
where $\sigma^\pm$ are the respective ladder operators. Herewith the Hamiltonian in Eq.~(\ref{Hamiltonian}) becomes
\begin{eqnarray}\label{Hamiltonian2}
  H^{(n)}\!\!&\!=\!&\! J\! \sin(\phi)  \! \sum_{j}\! \left( \! c_{j}^\dagger c_{j+1}^\dagger - c_{j}^\dagger c_{j+1}  + c_{j} c_{j+1}^\dagger - c_{j} c_{j+1}
  \right) \nonumber \\
  &\!+\!&\!J\!\cos(\phi) \sum_{j}\left( c_{j}^\dagger c_{j+n+1}
  - c_{j} c_{j+n+1}^\dagger\right.\nonumber\\
  && \;\;\;\;\;\;\;\;\;\;\;\;\;\;\;\;\;\;\;\; \left.+ c_{j}^\dagger c_{j+n+1}^\dagger - c_{j} c_{j+n+1}\right)
  \end{eqnarray}
One notes that herewith the cluster interaction is reduced from a $n+2$ interaction to a two-body interaction
between sites at distance $n+1$. After having reduced the problem to an effective two-body one the model can be diagonalized via the Fourier
transforms of the fermionic operators, i.e.
\begin{eqnarray}
\label{FT1}
 b_k & = & \frac{1}{\sqrt{N}} \sum_{j} c_k\; e^{-i\,kj}\; , \nonumber \\
 b_k^\dagger & = & \frac{1}{\sqrt{N}} \sum_{j} c_k^\dagger\; e^{i\,kj} \; ,
\end{eqnarray}
where the wave number $k$ is equal to $k=2 \pi l/N$ and $l$ runs from $-N/2$ to $N/2$ and $N$ is the total number of spins (sites) in the chain.
The Hamiltonian transforms to
\begin{eqnarray}
  \label{Hamiltonian3}
 H^{(n)} & = & \sum_{k>0} h_{k}^{(n)}
 \end{eqnarray}
 with
 \begin{eqnarray}
 h_{k}^{(n)}& =& 2\,i\, \delta_{k,n}\;  \left( b_{k}^\dagger b_{-k}^\dagger- b_{-k} b_{k}\right) \nonumber\\
 &&+
 2\, \varepsilon_{k,n}\; \left( b_{k}^\dagger b_{k}+ b_{-k}^\dagger b_{-k}-1 \right) \; , \nonumber
\end{eqnarray}
where the parameters $\delta_{k,n}$ and $\varepsilon_{k,n}$ are respectively given by
\begin{eqnarray}
  \label{deltaepsilon}
 \delta_{k,n}& = & J\;\left(\sin\left((n+1)k\right) \cos\phi+ \sin(k) \sin\phi\right)\;, \nonumber \\
 \varepsilon_{k,n} & = &  J\;\left(\cos\left((n+1)k\right) \cos\phi- \cos(k) \sin\phi\right)\; .
\end{eqnarray}
Via these transformations we re-wrote the Hamiltonian under investigation into the sum of non-interacting terms $h_{k}^{(n)}$, each one of them
acting only on fermionic states with wave number equal to $k$ or $-k$.  Each  $h_{k}^{(n)}$ corresponds to a four level system that can be
expressed in an occupation number basis by  $|1_k,1_{-k}\rangle$, $|0_k,0_{-k}\rangle$, $|1_k,0_{-k}\rangle$, $|0_k,1_{-k}\rangle$ and is,
explicitly, represented by the following matrix
\begin{equation}
 \label{Hamiltoniank}
h_{k}^{(n)}=\left(
\begin{array}{cccc}
2\; \varepsilon_{k,n} & +2\, i\; \delta_{k,n} & 0 & 0 \\
- 2\, i\; \delta_{k,n} & - 2\; \varepsilon_{k,n} & 0 & 0 \\
 0 & 0 & 0 & 0 \\
 0 & 0 & 0 & 0
\end{array}
\right) \; ,
\end{equation}
which ground state energy computes to
\begin{equation}
\label{Ek0}
 \!\!\!E_{k}^{(n)}\!\!=\!- 2\! \sqrt{\!\varepsilon_{k,n}^2\!+\!\delta_{k,n}^2}\!=\!- 2J\! \sqrt{\!1\!-\!\cos((n\!+\!2)k)\! \sin(2\phi)} \!\;.
\end{equation}
The associated ground state $|\psi_k^{(n)}\rangle$ is a superposition of $|1_k,1_{-k}\rangle$ and $|0_k,0_{-k}\rangle$
\begin{equation}
|\psi_{k}^{(n)}\rangle=\alpha_{k,n}\; |1_k,1_{-k}\rangle +\beta_{k,n}\; |0_k,0_{-k}\rangle
\end{equation}
with
\begin{eqnarray}
 \label{alphakbetak}
 \alpha_{k,n} &=& i \frac{\varepsilon_{k,n} +E_{k}^{(n)}}{\sqrt{\delta_{k,n}^2+(\varepsilon_{k,n}
 +E_{k}^{(n)})^2}} \; ,
 \nonumber \\
 \beta_{k,n} &=&  \frac{\delta_{k,n}}{\sqrt{\delta_{k,n}^2+(\varepsilon_{k,n}+ E_{k}^{(n)})^2}} \; .
\end{eqnarray}
Since the Hamiltonian is the sum of the non-interacting terms $h_{k}^{(n)}$, each one of them is acting on a different Hilbert space,
and the ground state of the total Hamiltonian is consequently a tensor product of all $|\psi_{k}^{(n)}\rangle$
\begin{equation}
 |\psi^{(n)}\rangle=\bigotimes_{k} |\psi_{k}^{(n)}\rangle\;.
\end{equation}
The associated energy density $E_{n,\phi}$ is the sum $E_k^{(n)}$ divided by the total number of the spins $N$. In the thermodynamic
limit the energy density becomes
\begin{equation}
 \label{E0}
E_{n,\phi}=-\frac{2J}{\pi} \int_0^\pi  \sqrt{1-\cos((n+2)k) \sin(2\phi)} dk \;.
\end{equation}

\begin{figure}
\includegraphics[width=0.49\textwidth]{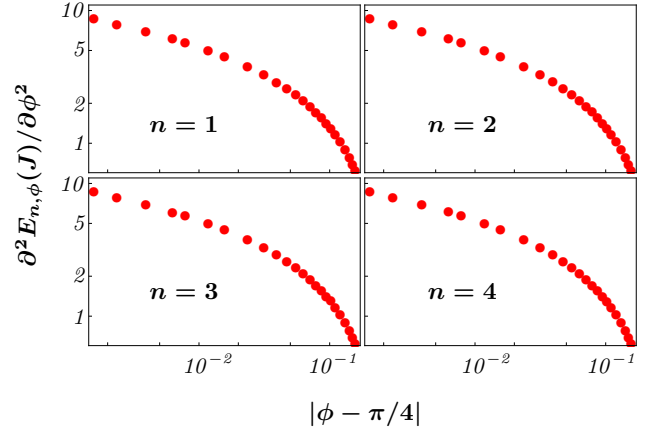}
\caption{(Color online) The graphes show the second derivative of the energy density of the ground state $E_{n,\phi}$ as a function of $\phi$ for
different cluster sizes $n+2$. The divergence is independent of $n$ at the critical value $\phi_c=\frac{\pi}{4}$ and corresponds to a vanishing
energy gap between the ground state and the first excited state.}
\label{secondderivativeenergy}
\end{figure}

According to the general theory of continuous phase transitions at zero temperature~\cite{S2011} the presence of a quantum critical point is
signaled by the divergence of the second derivative of the energy density with respect to the Hamiltonian parameter. In
Fig.~\ref{secondderivativeenergy} the second derivative of the energy density is plotted in dependence of $\phi$ and shows a divergence for the
value $\phi=\phi_c\equiv\pi/4$ independent of $n$.  The singularity is ultimately due to the vanishing of the energy gap between the ground state and the
first excited state at the critical value $\phi_c$ with the modes $k=\frac{j\pi}{n+2}$, where $j$ runs from $0$ to $n+1$ .

\section{The spin correlations functions}
\label{Spincorrelation}


To obtain a generic spin correlation function we can adapt the strategy that we used to compute the energy density. Then applying Wick's
theorem~\cite{W1950} simplifies the issue further since it makes it possible to express any multi-body fermionic correlation
function in terms of two-body correlation functions. More precisely, it is possible to prove~\cite{BM1971} that defining for each site $j$,
two fermionic operators, $A_j$ and $B_j$, via
\begin{equation}
 \label{A&B}
 A_j=c_j+c_j^\dagger \qquad\textrm{and}\quad  B_j=c_j-c_j^\dagger\;,
\end{equation}
any spin correlation function can be written as an ordered
product of these operators. Hence, due to Wick's theorem, any spin correlation function can be written as a
combination of one- and two-body expectation values involving only operators $A_j$ and $B_k$ on the same or different sites.
With Eq.~(\ref{alphakbetak}) we obtain
\begin{eqnarray}
\label{correlations}
 \langle A_i \rangle & = &  0\;, \nonumber \\
 \langle B_i \rangle & = &  0\;, \nonumber \\
 \langle A_i A_k\rangle  & = & \delta_{ik}\;, \nonumber \\
 \langle B_i B_k\rangle  & = & -\delta_{ik}\;,  \\
 \langle B_i A_k\rangle  & = & G_{i,k}(n,\phi)\;. \nonumber
\end{eqnarray}
The fact that we have that both $\langle A_i \rangle = \langle B_i \rangle=0$ and  $\langle A_i A_k\rangle=\langle B_i B_k\rangle =0$
for $i \neq k$ has several important consequences.
In fact, let us consider a spin correlation function associated with an operator that is the product of many local spin operators, each one acting
onto different spins, in which $\sigma_j^x$ and/or $\sigma_j^y$  appears an odd number of times on different sites. To this operator we may
associate a fermionic operator made by a different number of $A_j$ and $B_j$ operators acting onto different spins. Therefore, when we apply the
Wick's theorem,
we have an expectation value of a single fermionic operator and/or an expectation value of two operators of the same kind onto different
spins. Hence, taking into account Eq.~(\ref{correlations}), such spin correlation functions have to vanish.
Consequently, the only correlation function that can be different
from zero are the ones associated with an operator that is a product of local spin operators in which both $\sigma_j^x$ and $\sigma_j^y$ appear an
even number of times.

To obtain the explicit expression of the non-zero spin correlation functions we need to evaluate $G_{i,k}(n,\phi)$. At first we note
that, in the thermodynamic limit, the $G_{i,k}(n,\phi)$ must be independent from the choice of $i$ and $k$ but may depend on
their relative distance \mbox{$r=i-k$}. With eq.~(\ref{alphakbetak}) we find $G_{i,k}(n,\phi)=G_{r}(n,\phi)$ with
\begin{equation}
 \!\!G_r(n,\phi)\!=\!\frac{1}{\pi}\! \int_0^\pi\!\!
\frac{\cos(\!k(\!n\!+\!1\!+\!r\!)\!)\!\cos\phi\!-\!\cos(\!k(\!r\!-\!1\!)\!) \sin\phi}{\sqrt{1-\cos((n+2)k)\! \sin(2\phi)}} dk \; .
\end{equation}
Solving this integral we find that if $r\neq l(n+2)+1$, where $l$ is an integer number that runs from $-\infty$ to $\infty$,
then the $G_{r}(n,\phi)$ vanishes for all values of $\phi$. This fact, as we show in Sec.~\ref{Entanglementproperties}, plays a fundamental
role in the behavior of the entanglement property among different spins.

Obviously, from Eq.~(\ref{correlations}) and the explicit expressions $G_{i,k}(n,\phi)$ one can recover all spin correlation functions of interest.
Here we wish to point out some interesting results about some specific ones.

If one allows for a magnetization along the $z$ direction, i.e. $\langle \sigma_j^z \rangle$, one finds that it equals $G_{0}(n,\phi)$ and,
therefore, vanishes identically for all possible values of $\phi$ and $n$. Let us consider two-body spin correlation functions that can be written
as $\langle \sigma_i^\mu \sigma_{i+r}^\mu\rangle$ with $\mu=x,y,z$.
If  $\mu$ coincide with $z$ the correlation function can be written as
\begin{equation}
 \langle \sigma_i^z \sigma_{i+r}^z\rangle= G_{0}(n,\phi)-G_{r}(n,\phi)G_{-r}(n,\phi) \;.
\end{equation}
Since $G_{r}(n,\phi)$ with $r \neq l(2+n)+1$ vanishes we find that
\begin{equation}
 \label{zcorr}
 \langle \sigma_i^z \sigma_{i+r}^z\rangle=0
\end{equation}
for all values of $n$ and $\phi$. Setting $\mu=x,y$ the spin correlation functions are given by the determinant
\begin{equation}
 \label{correlationmatrixx}
\!\! \!\!\!\!\langle \sigma_i^x \sigma_{i+r}^x\rangle\!=\!
 \left|
 \begin{array}{cccc}
 \! G_{-1}(n,\phi) & G_{-2}(n,\phi) &\! \cdots \!& \!\!G_{-r}(n,\phi)\! \\
 \! G_{-2}(n,\phi) & G_{-1}(n,\phi) & \!\cdots \!& G_{-r+1}(n,\phi)\!\! \\
 \! \vdots & \vdots & \!\ddots \!& \vdots\! \\
 \! G_{-r}(n,\phi) &\!\! G_{-r+1}(n,\phi) \!\!& \!\cdots\! &\!\! G_{-1}(n,\phi)\!
 \end{array}
\right|\;,
\end{equation}
\begin{equation}
 \label{correlationmatriyy}
\!\! \!\!\!\!\langle \sigma_i^y \sigma_{i+r}^y\rangle\!=\!
 \left|
 \begin{array}{cccc}
 \! G_{1}(n,\phi) & G_{2}(n,\phi) &\! \cdots \!& \!\!G_{r}(n,\phi)\! \\
 \! G_{2}(n,\phi) & G_{1}(n,\phi) & \!\cdots \!& \!\!G_{r-1}(n,\phi)\! \\
 \! \vdots & \vdots & \!\ddots \!& \vdots\! \\
 \! G_{r}(n,\phi) & G_{r-1}(n,\phi) & \!\cdots\! & \!\!G_{1}(n,\phi)\!
 \end{array}
\right|\;.
\end{equation}
Numerical evaluations  reveal that $\langle \sigma_i^x \sigma_{i+r}^x\rangle$
is non-vanishing only when $r$ is an integer multiple of $n+2$,
in strong contrast to the correlation function  $\langle \sigma_i^y \sigma_{i+r}^y\rangle$, which is always non-zero. It changes from negative to
positive values in the case $r$ varies from odd to even values. This is expected due to the anti-ferromagnetic nature of the Ising interaction.

\section{The order parameters}
\label{Orderparameters}

As we have seen in Sec.~\ref{Solution}, the behavior of the second derivative of the ground state energy density shows a phase
transition at $\phi=\phi_c\equiv \pi/4$ for all $n$. Now we characterize the properties of these two phases via the help of the spin correlation
functions (Sec.~\ref{Spincorrelation}).

Let us start from the phase $\phi>\phi_c$, i.e. when the system is dominated by a two-body anti-ferromagnetic Ising interaction along the $y$ spin
direction.  Due to the $Z_2$ symmetry of the Hamiltonian~(\ref{Hamiltonian}) we cannot compute the staggered magnetization by directly applying the
definition
\mbox{ $ m_y= (-1)^j \langle \sigma_j^y \rangle$} since this gives always a vanishing result. Approaching the problem we may first evaluate the
value of the magnetization with respect to its relation to the long distance correlation function along the same spin direction, i.e.
\begin{equation}
 \label{staggeredmagnetization}
 m_y=\sqrt{\lim_{r\rightarrow \infty} (-1)^r \langle \sigma_i^y \sigma_{i+r}^y\rangle}\;.
\end{equation}
This can be evaluated via the help of Eq.~(\ref{correlationmatriyy}). We have computed for different $n$ numerically the quantity
$(-1)^r \langle \sigma_i^y \sigma_{i+r}^y\rangle$ with $r$ up to 200, showing that an increase of the distance $r$ results only in a very small
variation of $m_y$ (of a factor less than $10^{-8}$) for each value of $\phi>\frac{\pi}{4}$.

\begin{figure}[t]
\includegraphics[width=0.49\textwidth]{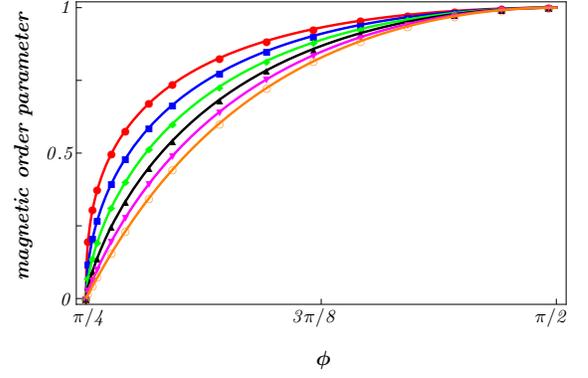}
\caption{(Color online) Behavior of magnetic order parameter in the Ising phase $\phi>\phi_c$ plotted for $n=1,2,\dots,6$: red (uppermost curve)
$n=1$; blue $n=2$;
green $n=3$; black $n=4$; magenta $n=5$; orange (lowest curve) $n=6$ . The dots represent the numerical
results of $m_y$ for $r$ going to infinity,  Eq.~(\ref{staggeredmagnetization}), whereas the curves corresponds to the guessed function of the
staggered magnetization $m_y^{(n)}$, Eq.(\ref{staggeredmagnetizationformula}).}
\label{staggeredmagnetizationfig}
\end{figure}

The results that we have obtained for different $n$ and $\phi$ are plotted in Fig.~(\ref{staggeredmagnetizationfig}). This shows the presence of an
anti-ferromagnetic phase along the $y$-direction for $\phi>\phi_c$ independent of the value of $n$. However, differently from what happens for the
second derivative of the density of the ground state energy, the staggered magnetization shows a clear dependence on $n$. Analyzing the numerical
data we can conclude that the staggered magnetization has the following dependence on $\phi\ge \phi_c$ and $n$:
\begin{equation}
 \label{staggeredmagnetizationformula}
 m_y^{(n)}=\left(1-\tan(\phi)^{-2}\right)^{\frac{n+2}{8}} \; .
\end{equation}
From that we can deduce the critical exponent $\beta$ over $n$
\begin{equation}
 \label{beta}
 \beta(n)=\frac{n+2}{8}\;.
\end{equation}
The fact that the critical exponent $\beta$ depends on $n$ means that the class of symmetry to which the models given by the
Hamiltonian~(\ref{Hamiltonian}) belongs depends on $n$.

The situation changes drastically when we move in the phase below the quantum critical point $\phi<\phi_c$. In this phase our Hamiltonian is
dominated by the many-body interaction terms and $m_y$ drops to zero for any $n$. It is not straightforward to find a proper candidate or the
role of order parameter as it was for the anti-ferromagnetic phase discussed above. However, after an heavy numerical analysis we were able to
obtain a clear picture on the ongoing physics of the system.

\begin{figure}[t]
\includegraphics[width=0.49\textwidth]{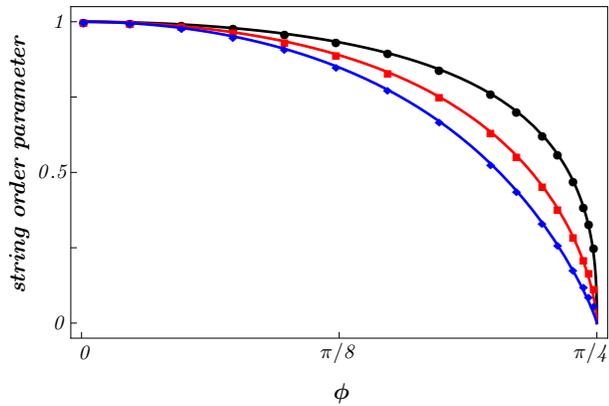}
\caption{(Color online) Behavior of the string order parameter for $\phi<\phi_c$ for $n=1,3,5$: black (upper curve) $n=1$;  red (middle curve) $n=3$;
blue (lower curve) $n=5$. The dots represent the numerical
results of the string order parameter $S_n$ given in Eq.~(\ref{stringorderparameter}), whereas the curves correspond to the behavior of the string
order parameter $S^{(n)}$ defined in Eq.~(\ref{BneOn}).}
\label{stringorderparameterfig}
\end{figure}

For a system with odd $n$ we can define a string order parameter as
\begin{eqnarray}
\label{stringorderparameter}
 S_n&=&\sqrt{\lim_{r \rightarrow \infty} \langle \sigma_1^x \sigma_2^y \sigma_3^x\! \cdots\! \sigma_{n+1}^y \mathcal{O}  \sigma_{r-n+1}^y\! \cdots\!
 \sigma_{r-2}^x \sigma_{r-1}^y \sigma_{r}^x\rangle}\;,\nonumber\\
\end{eqnarray}
where the operator $\mathcal{O}=O^z_{n+1,r-2(n+1)}$. In Fig.~\ref{stringorderparameterfig} the behavior of this string order parameter $S_n$ for
$n=1,3,5$ is plotted. The existence of such a non-vanishing string order parameter can be traced back to the presence of diverging localizable
entanglement~\cite{PVMC2005,CR2005}. This signals the presence of a symmetry protected topological order.

\begin{figure}[t]
\includegraphics[width=0.49\textwidth]{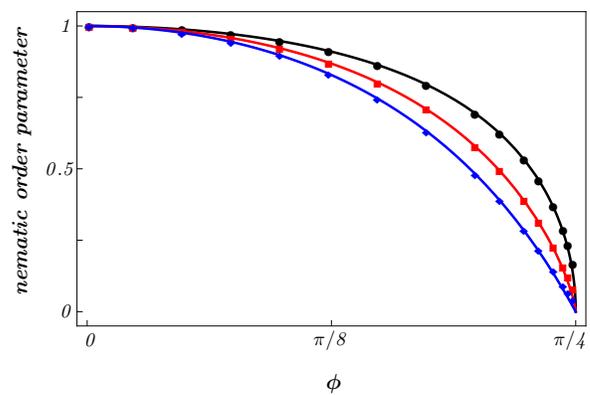}
\caption{(Colour online) Behavior of the nematic order parameter for \mbox{$\phi<\phi_c$} for $n=2,4,6$; black (upper curve) $n=2$;  red (middle curve) $n=4$;
blue (lower curve) $n=6$. The dots represent the
numerical results of the nematic order parameter $B_n$, Eq.~(\ref{blockorderparameter2}), whereas the curves correspond to nematic order
parameter $B^{(n)}$, Eq.~(\ref{BneOn}).}
\label{blockmagnetizationfig}
\end{figure}

For even $n$ we find that the phase is a nematic one thus we can define the following order parameter (since the system is translation invariant
the quantity is understood to not depend on the particular $i$)
\begin{equation}
\label{blockorderparameter}
\mathcal{B}_{n}=\langle \mathcal{O}_{i,n}\rangle=\langle \sigma_i^x \sigma_{i+1}^y \sigma_{i+2}^x\! \cdots\! \sigma_{i+n}^x  \rangle
\end{equation}
As in the staggered magnetic order phase $\mathcal{B}_n$ cannot be evaluated directly since it vanishes (for any even $n$ the operators
$\sigma_i^x$ or $\sigma_i^y$ appear an odd number of times). Again we can circumvent this problem by defining
\begin{equation}
\label{blockorderparameter2}
B_n=\sqrt{\lim_{r \rightarrow \infty }\langle \mathcal{O}_{i,n} \mathcal{O}_{i+r,n} \rangle}\;.
\end{equation}
In Fig.~\ref{blockmagnetizationfig} we plotted the behavior of $B_n$ for $n=2,4,6$.

Analyzing the numerical data obtained for both defined string order parameters, $S_n$ and $B_n$, we find finally the same dependence on $n$
and $\phi$, i.e.
\begin{eqnarray}\label{BneOn}
 S^{(n)}&=&\left(1-\tan(\phi)^{2}\right)^{\frac{n+2}{8}}  \nonumber \\
 B^{(n)}&=&\left(1-\tan(\phi)^{2}\right)^{\frac{n+2}{8}} \; .
\end{eqnarray}

Summarizing all results we can formulate a general concise formula for all order parameters of the whole class of models given by the
Hamiltonian~(\ref{Hamiltonian}):
\begin{eqnarray}\label{generalOrderPar}
\textrm{Order Parameter}&=& \left(1-\tan(\phi)^{-2 sgn(\phi-\frac{\pi}{4})}\right)^{\frac{n+2}{8}}\;.\nonumber\\
\end{eqnarray}
Moreover, the existence of a duality, i.e. a transformation that brings the order parameters before and after the critical point in relation, is
thus proven.

In summary, we find that for both phases we can define order parameters that each is ruled by the dominated interactions, i.e. Ising interaction or
multi-body cluster interaction. In the multi-body cluster phase a strong dependence on the size of the cluster $n+2$ is present revealing either a
nematic phase (even $n$) or a topologically ordered phase (odd $n$).

\section{The entanglement properties}
\label{Entanglementproperties}

In this section we analyze the entanglement properties between adjacent spins as well as between a block of spins and the remaining part of the
chain. Despite the complexity of the class of models under investigation we obtain general results showing the relevance of the entanglement
features in these complex matter systems.

The object of matter is the reduced density of $m$ spins, which is obtained by taking the trace over all remaining spins of the ground state.
Any such reduced density matrix we can decompose by the spin correlation functions
\begin{equation}
 \label{densitymatrix}
 \rho_m^{(n)}=\frac{1}{2^m} \sum_{\alpha_1,\dots,\alpha_m} \langle \sigma_1^{\alpha_1}\sigma_2^{\alpha_2}\cdots\sigma_m^{\alpha_m}\rangle
 \sigma_1^{\alpha_1}\sigma_2^{\alpha_2}\cdots\sigma_k^{\alpha_k}\; ,
\end{equation}
where $\alpha_i$ runs from $0,x,y,z$ and $\sigma_i^{0}$ denotes the identity.

The next subsection introduces the concept of different types of multipartite entanglement. Then we compute the entanglement properties of
adjacent spins and the entanglement between a block of spins and the remaining part of the chain.

\subsection{Definition of hierarchies of multipartite separability}

The quantum separability problem reduces for bipartite entangled systems to the question of whether the state is entangled or not. In the
multi-partite case the problem is more involved. First, there exist different hierarchies of separability since an $n$-partite entangled
state $\rho$ may be a convex combination of pure entangled states with maximally $k$ entangled particles. Any tripartite pure state, e.g., can be
written as
\begin{eqnarray}
|\psi_{k=3}\rangle&=&|\phi_A\rangle\otimes|\phi_B\rangle\otimes|\phi_C\rangle\nonumber\\
|\psi_{k=2}\rangle&=&|\phi_A\rangle\otimes|\phi_{BC}\rangle,\quad|\phi_B\rangle\otimes|\phi_{AC}\rangle\nonumber\\
&&\textrm{or}\quad |\phi_{AB}\rangle\otimes|\phi_{C}\rangle\nonumber\\
|\psi_{k=1}\rangle &=&|\psi\rangle_{ABC}
\end{eqnarray}
where $k$ gives the number of partitions dubbed the $k$-separability. In general a pure state $\ket{\Psi^k}$ is called $k$-separable, if and only if
it can be written as a tensor product of $k$ factors $\ket{\psi_i}$, each of which describes one or several subsystems, i.e.
\begin{eqnarray}\label{k-separability}
\ket{\Psi^k}&=&|\psi_1\rangle\otimes\ket{\psi_2}\otimes\dots\otimes\ket{\psi_k}\;=\;|\psi_1\psi_2\dots\psi_k\rangle \; .\end{eqnarray}
A mixed state $\rho$ is called $k$-separable, if and only if it can be decomposed into a mixture of $k$-separable pure states
\begin{eqnarray} \rho = \sum_i p_i \ket{\Psi_i^k}\bra{\Psi_i^k}
\end{eqnarray}
where all $\ket{\Psi_i^k}$ are $k$-separable (possibly with respect to different $k$-partitions) and the $p_i$ form a probability distribution.
An $n$-partite state (pure or mixed) is called fully separable if and only if it is $n$-separable. It is called \textit{genuinely multi-partite
entangled} if and only if it is not bi-separable (2-separable). If neither of these is the case, the state is called \textit{partially}
multipartite entangled or \textit{partially} multipartite separable. Note that obviously a $k=3$-separable state is necessarily also
$k=2$-separable, thus $k$-separable states have a nested-convex structure.

In particular, note that the following tripartite mixed state
\begin{eqnarray}
\rho&=&\;\sum_i p_i\; |\psi_i\rangle_{AB}\langle\psi_i|_{AB}\otimes|\sigma_i\rangle_C\langle \sigma_i|_C\nonumber\\&& +\sum_i q_i\;
|\chi_i\rangle_{AC}\langle\chi_i|_{AC}\otimes|\tau_i\rangle_B\langle \tau_i|_B\nonumber\\
&&+\sum_i r_i\; |\xi_i\rangle_{BC}\langle\xi_i|_{BC}\otimes|\omega_i\rangle_A\langle \omega_i|_A
\end{eqnarray}
with $p_i,q_i,r_i\geq 0$ and $\sum p_i+q_i+r_i=1$ is bi-separable though it is not bi-separable with respect to a certain splitting. This property
and the fact that the convex sum of pure states is not unique are the reasons why it is hard to detect genuine multipartite entanglement, i.e. a
state that cannot be written in the above form. Consequently, the entanglement characterization of multi-partite states needs more than the
combination of bipartite entanglement criteria~\cite{HMGHframework}.

\subsection{Entanglement properties among adjacent spins}

Let us start by analyzing the case of $m$ adjacent spins thus having a maximum distance of $r=m-1$. Then all spin correlation functions can be
expressed by $G_{r}(n,\phi)$ with $-(m-1)<r<m-1$.

\begin{figure}[t]
\includegraphics[width=0.49\textwidth]{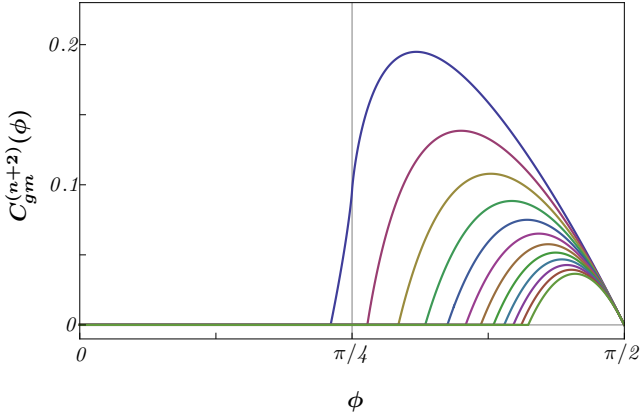}
\caption{(Color online) Dependence of the genuine multipartite concurrence $C_{gm}^{(n+2)}$ as function of the weighted interactions $\phi$ for
different $n$ that runs from $1$ (highest curve) to $12$ (lowest curve). Note that only for $n=1$ genuine tripartite entanglement is non-zero
before and after the critical point and, generally, genuine $n+2$-partite entanglement decrease with increasing cluster size.}
\label{multipartiteentanglementfig}
\end{figure}

\begin{theorem}
If the number of adjacent spins $m$ is smaller than the cluster size, i.e. $m<n+2$, then all $k\leq m$-partite entanglement vanishes, i.e. the
reduced state is $k\leq m$-separable. If the number of adjacent spins equals the cluster size, i.e. $k=n+2$, then there exists a finite range of
values of $\phi$ for which the reduced density matrix to this set of
spins is genuinely $n+2$-partite entangled (plotted in Fig.~\ref{multipartiteentanglementfig}).
\end{theorem}

\textbf{Proof:} Let us start with $m<n+2$. In Sec.~\ref{Spincorrelation} we have computed all spin correlation functions $G_r(n,\phi)$ and found
that they vanish if $r\not=(n+2)l+1$. The reduced density matrix $\rho_m^{(n)}$ depends only on a single function, i.e. $G_1(n,\phi)$. This implies
that only spin correlation functions that are different from
zero are the ones along the $y$-direction. Consequently, the reduced matrix $\rho_m^{(n)}$ is a mixture of states being eigenvectors to the
single-spin operators $\sigma_j^y$. Applying the following local unitary operators
\begin{equation}
 \label{Unitary}
 U_j=\exp\left(-i \frac{\pi}{4} \sigma_j^x\right) \; ,
\end{equation}
brings the density matrix $\rho_m^{(n)}$ of $m$ adjacent spins into a diagonal form which obviously is separable.

In the case where the adjacent spins equals the cluster size, $m=n+2$, the reduced density matrix $\rho_{n+2}^{(n)}$ depends on $G_1(n,\phi)$ and
$G_{-(n+1)}(n,\phi)$ that corresponds to the spin cluster correlation function $\langle \sigma_k^x O_{k,n}^z\sigma_{k+n+1}^x \rangle$. Again
applying the above defined local unitary operators $U_j$ to each spin we obtain a reduced density matrix $\rho_m^{(n)}$ that has an
X-form~\cite{YE2007}, i.e. only entries on both diagonals are nonzero. It has been shown that if a density matrix can be written in such an
$X$-form, the genuine multipartite entanglement can be exactly evaluated by a certain measure, dubbed genuine multipartite concurrence introduced
in Refs.~\cite{HH2008,MCCSGH2011,HHBE2012}. Thus, by applying the above defined local unitaries we find the following expression for the genuine
$(n+2)$-partite concurrence for any $n$ and $\phi$
\begin{eqnarray}\label{genuinemltipartiteconcurrence}
\lefteqn{C_{gm}^{(n+2)}(\phi)\;=}\nonumber\\
&&\max[0,\frac{1}{2^{n+1}} \left(1-G_{1}(n,\phi)\right)^{n+1}\cdot \left(G_{-(n+1)}(n,\phi) +1\right)\nonumber\\
&&\qquad\qquad\qquad -1 ]\;.
\end{eqnarray}

In Fig.~\ref{multipartiteentanglementfig} we have plotted the genuine $(n+2)$-partite concurrence for $n=1,\dots,12$ and find for certain $\phi$
non-zero values. Q.E.D.\\
\\
Looking more carefully at the curves, one observes a similar behavior for all $n$ and, except for $n=1$, a non-zero value of genuine
$(n+2)$-partite multipartite is only obtained in the Ising phase. Moreover, the genuine $(n+2)$-partite concurrence is always smaller for bigger
cluster sizes. That proves that the entanglement in the ground state becomes robust against the Ising-interaction (remember the ground state of
$\phi=0$ is a graph state and for $\phi=\frac{\pi}{2}$ a totally factorized state~\cite{GAI2008,GAI2009,GAI2010}). Consequently, higher cluster
sizes allow for better properties for running quantum algorithms.

\begin{figure}[t]
\includegraphics[width=0.49\textwidth]{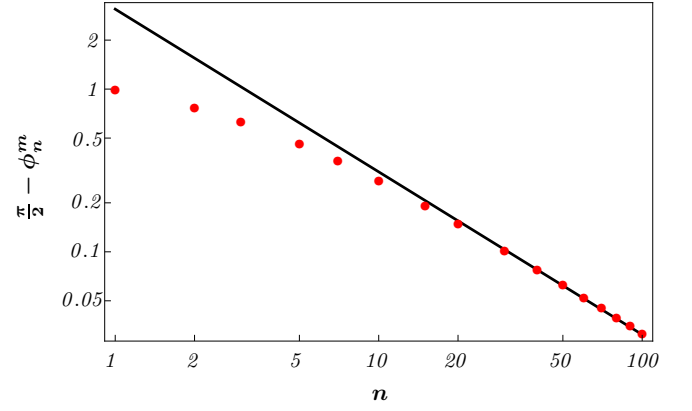}
\caption{(Color online) Behavior of $\frac{\pi}{2}-\phi_\textrm{max}^{(n)}$ as function of $n$ where the maximization is taken for the genuine
multipartite concurrence. The red dots are the result of the numerical maximization
for any $n$ while the black line represent the fit, obtained for large $n$, in  Eq.~(\ref{powerlawmax}).}
\label{multipartiteentanglementpositionfig}
\end{figure}

\begin{figure}[t]
\includegraphics[width=0.49\textwidth]{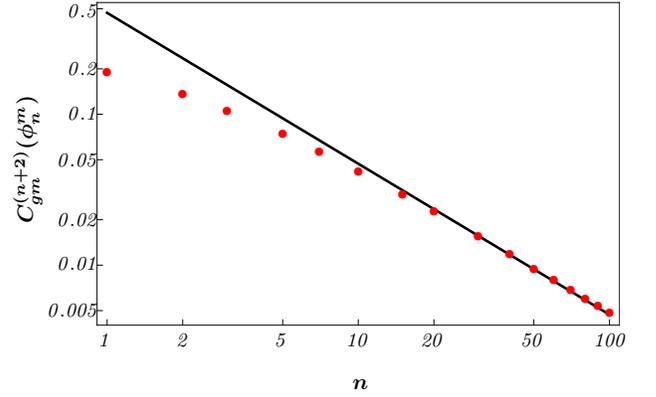}
\caption{(Color online) Dependence of the maximum value of the genuine multipartite concurrence $C_{gm}^{(n+2)}(\phi_\textrm{max}^{(n)})$ on the
cluster size $n+2$. The (red) dots are the results of the numerical maximization for any $n$ whereas the (black) line represent the fit obtained
for large $n$ presented in Eq.~(\ref{powerlawmax}).}
\label{multipartiteentanglementvaluefig}
\end{figure}

\begin{figure}[t]
\includegraphics[width=0.49\textwidth]{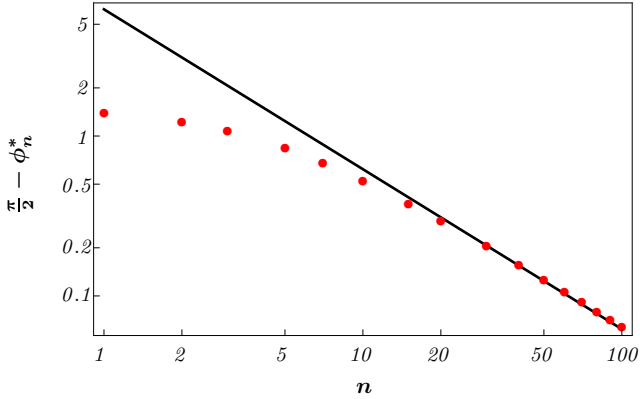}
\caption{(Color online) Dependence of $\frac{\pi}{2}-\phi_*^{(n)}$ as function of $n$. The (red) dots are the result of the numerical result
for specific cluster sizes $n+2$ whereas the (black) line represents the fit result obtained for large $n$ presented in Eq.~(\ref{powerlawmax1}).}
\label{multipartiteentanglementzerofig}
\end{figure}
In Fig.~\ref{multipartiteentanglementpositionfig} and Fig.~\ref{multipartiteentanglementvaluefig} a deeper analysis of the entanglement properties
of the reduced matrix can be found concerning the maximal value of the weight $\phi_\textrm{max}^{(n)}$ of the two interactions which corresponds to
the maximal reachable value of genuine $n+2$-partite concurrence $C_{gm}^{(n+2)}$. Both values show a similar dependence that for $n>10$ is in good
approximation given by
\begin{eqnarray}
 \label{powerlawmax}
\phi_\textrm{max}^{(n)}&=& \frac{\pi}{2}-\frac{3.1}{n} \nonumber \\
 C_{gm}^{(n+2)}(\phi_\textrm{max}^{(n)})&=&\frac{0.47}{n} \; ,
\end{eqnarray}
where the numerical coefficients are obtained by a best fit algorithm. Analogously, the point in which the genuine $(n+2)$-partite
entanglement becomes different from zero depends on the inverse of $n$, plotted in Fig.~(\ref{multipartiteentanglementzerofig}),
\begin{equation}
 \label{powerlawmax1}
 \phi_*^{(n)}= \frac{\pi}{2}-\frac{6.2}{n} \;.
\end{equation}
From these equations we immediately reveal an interesting relation between $\phi_\textrm{max}^{(n)}$ and $ \phi_*^{(n)}$, i.e.
\begin{equation}
 \label{powerlawmaxrelation}
 \left( \phi_*^{(n)}- \frac{\pi}{2}\right)=2 \left(\phi_\textrm{max}^{(n)}- \frac{\pi}{2}\right) \;,
\end{equation}
valid for large $n$.

In summary, these cluster-Ising models with different cluster sizes have interesting local entanglement properties. There is no bipartite,
tripartite,..., $n+1$ entanglement, but only for large enough values of Ising interaction $\phi>\frac{\pi}{4}$ one finds local entanglement,
in particular only genuine $n+2$-partite multipartite entanglement. In Ref.~\cite{HKN2006} the authors computed that the maximal value of maximal
possible entanglement of two adjacent spins in a translation invariant chain was found to give a (bipartite) concurrence of $C=0.434467$. This
optimal value serves for interpreting entanglement
values obtained for real physical systems. In the very same manner the maximal values of the multipartite entanglement quantified by the above
introduced genuine multipartite entanglement measure serves as an reference for real physical system exhibiting cluster and Ising interactions.

\subsection{Entanglement properties between a block of spins and the rest of the chain}

Another important property to analyze in multipartite systems concerns the entanglement features of a block of $m$ spins with the rest of the chain
and how it classifies to the holomorphic and anti-holomorphic sectors in conformal field theories.

For that purpose we have to compute the von Neumann entropy of the reduced density matrix of $m$ spins,
\begin{equation}
\label{VNEdefinition}
 S_{m}^{(n)}=Tr(\rho_m^{(n)} \log_2(\rho_m^{(n)}))\;.
\end{equation}
Using the methods developed in Ref.~\cite{VLRK2003,LRV2004} we find
\begin{equation}
 \label{VNEdefinition1}
 S_{m}^{(n)} = \sum_{j=1}^m H_{\textrm{Shannon}}\left(\frac{1+\nu_j}{2}\right)
\end{equation}
where $H_{\textrm{Shannon}}(x)$ is the Shannon entropy
\begin{equation}
 \label{Shannonentropy}
 H_{\textrm{Shannon}}(x)= -x \log_2(x) -(1-x) \log_2(1-x)\; ,
\end{equation}
and $\nu_j$ is the imaginary part of the eigenvalues of the matrix
\begin{equation}
\label{gammaprime}
 \Gamma'= \delta_{ij}-i \Gamma_m
\end{equation}
with
\begin{equation}
\label{gamma}
 \Gamma_m=\left( \begin{array}{cccc}
 \! \Pi_0 & \Pi_{-1} &\! \cdots \!&  \Pi_{-m+1} \! \\
 \! \Pi_{1} & \Pi_0 & \!\cdots \!&  \Pi_{-m+2}\! \\
 \! \vdots & \vdots & \!\ddots \!& \vdots\! \\
 \! \Pi_{m-1} &\Pi_{m-2} & \!\cdots\! & \Pi_0\!
 \end{array}
\right)
\end{equation}
and
\begin{equation}
\label{pi}
 \Pi_r=\left(
 \begin{array}{cc}
 0 & G_{r}(n,\phi) \\
 -G_{-r}(n,\phi) & 0 \\
 \end{array}
\right)\;.
\end{equation}

\begin{figure}[t]
\includegraphics[width=0.49\textwidth]{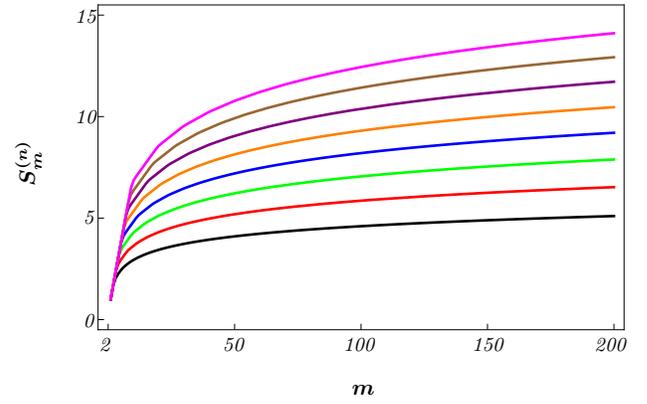}
\caption{(Color online) Here the von Neumann entropy $S^{(n)}_m$, Eq.~(\ref{VNEdefinition}), in dependence of a block size $m$ for different
cluster sizes $n$ is plotted from which we fit the numerical solution of the von Neuman entropy, see Eq.~(\ref{numericalfitvonneumann}). The value
of $n$ runs from $1$ lowest (black) curve to $8$ the highest (pink) curve.}
\label{voneumannfig}
\end{figure}

We have evaluated numerically the von Neumann entropy for blocks of length ranging from 2 to 200 spins at the critical point $\phi_c$ for
$n$ that runs from $1$ to $10$. The obtained values of the von Neumann entropy are displayed in Fig.~\ref{voneumannfig}.

Analyzing the numerical data we deduce
\begin{equation}
\label{numericalfitvonneumann}
 S_{m}^{(n)}\simeq(0.32+0.18\;n)\; \log_2 m+const(n)
\end{equation}
The multiplicative constant in front of the logarithmic term is known to be related to the central charge of the $1+1$ dimensional conformal
theory describing the critical behavior of
the chain via the relation~\cite{HLW1994}
\begin{equation}
 \label{cft}
 S_m = \frac{\mathsf{c}+\overline{\mathsf{c}}}{6}\; \log_2 m\;,
\end{equation}
where $\mathsf{c}$ and $\overline{\mathsf{c}}$ are the central charges of the so-called holomorphic and anti-holomorphic sectors of the conformal
field theory. Due to the existence of a duality in the system under investigation we have that $\mathsf{c}=\overline{\mathsf{c}}$ and hence, via
Eq.~(\ref{numericalfitvonneumann}) we
obtain
\begin{equation}
\label{centralcharge}
  \mathsf{c}\simeq 3\cdot (0.32+0.18\;n)\;.
\end{equation}
For two quantum one-dimensional systems to belong to the same universality class they need to have the same central charge. Since in our case we
find a dependence on $n$, this central charge, in addition to the critical exponent $\beta$, Eq.~(\ref{beta}), of the order parameters,
Eq.~(\ref{generalOrderPar}),
proves that the many-body cluster-Ising models fall into different classes with respect to their symmetries.

\section{Conclusions}
\label{Conclusions}

In summary, we have analytically solved, characterized and analyzed the properties of a family of models that we named $n$-cluster-Ising models.
These are models characterized by different cluster sizes ($n+2$) and different weighted cluster interaction and Ising interaction. We proved that
there occurs a phase transition exactly when both interactions are equally weighted and, interestingly, independent of the cluster size.

With respect to their symmetries the family of models falls into different classes proved via the dependence on $n$ of the critical exponent
$\beta$ of properly defined order parameters and the central charge of the holomorphic and anti-holomorphic sectors in conformal field theories.
In particular we find that the cluster phase has very different orderings for odd or even cluster size, namely a topological or a nematic order.
Since nematic order usually shows up only for non-analytically solvable systems these cluster-Ising models may become a prototype testing model for
exploiting the physical potential of nematic ordering of spins.

In the next step we have investigated how the apparent complexity of the ordering translates to the multipartite entanglement properties shared
among spins or block of spins with the rest of the system. Surprisingly, exactly all reduced density matrices with $m$ adjacent spins smaller than
the cluster size ($=\;n+2$ adjacent spins) posses no entanglement, whereas the reduced density matrices for exactly the cluster size ($n+2$)
possesses \textit{genuine} $n+2$-multipartite entanglement if the Ising interaction is strong enough, but not maximal
(see Fig.~\ref{multipartiteentanglementfig}). This absence of bipartite or $n-1$-partite multipartite entanglement is very different from other one
dimensional spin models, i.e. as the Ising one~\cite{OAFF2002,ON2002} or the $XY$-model~\cite{GH2013}. That computation was possible,
because the symmetries of the Hamiltonian constrain the state space in the Hilbert-space in a non-trivial way enabling even the computation of a
measure of genuine multipartite entanglement. From the quantum information perspective these results show that increasing the cluster size reduces
local entanglement and, herewith, the robustness of the performance of any quantum algorithm. From the perspective of comparison of different
condensed matter systems the family of models serves as a reference system of the possible amount of local genuine multipartite entanglement that
can be shared.

Our family of models can be generalized with respect to higher dimensions both in space and degrees of freedom (higher spins). These models may
become a good testing ground for non-trivial spin orderings and serve as a prototype for studying the potential of a quantum computer.

\section*{Acknowledgments} The authors
acknowledge gratefully the Austrian Science Fund (FWF-P23627-N16). We thank Benjamin Rogers for carefully reading the manuscript.

\end{document}